\documentclass[aps,pre,twocolumn,superscriptaddress]{revtex4-1}

\usepackage{amsmath,amssymb,amsfonts}
\usepackage{color}
\usepackage{varioref}
\usepackage{wrapfig}
\usepackage[FIGTOPCAP, hang, raggedright, nooneline]{subfigure}
 \usepackage{subfigmat}
\usepackage{graphicx}
\usepackage{dcolumn}
\usepackage{bm}

   \newcommand{\JCP}{J. Comp. Physics}

     \newcommand{\PRA}{Phys. Rev. A}
  \newcommand{\PRE}{Phys. Rev. E} \newcommand{\PRL}{Phys. Rev. Lett.}
  \newcommand{\PF}{Phys. Fluids}

\begin{document}


\title{Analysis of Boundary Slip in a Flow with an Oscillating Wall}



\author{Joseph John Thalakkottor}
\email[]{tjjoseph@ufl.edu}
\affiliation{Department of Mechanical and Aerospace Engineering, University of Florida.}
\author{Kamran Mohseni}%
\email[]{mohseni@ufl.edu}
\thanks{Department of Mechanical and Aerospace Engineering, University of Florida, Gainesville, FL-32611}
\affiliation{Department of Mechanical and Aerospace Engineering, University of Florida.}
\affiliation{Department of Electrical and Computer Engineering, University of Florida.}


\date{\today}

\begin{abstract}
 Molecular dynamic (MD) simulation is used to study slip at the fluid-solid boundary in an unsteady flow based on the Stokes second problem. An increase in slip is observed in comparison to the steady flow for shear rates below the critical shear rate of the corresponding steady flow. This increased slip is attributed to fluid inertial forces not represented in a steady flow. An unsteady mathematical model for slip is established, which estimates the increment in slip at the boundary. The model shows that slip is also dependent on acceleration in addition to the shear rate of fluid at the wall. By writing acceleration in terms of shear rate, it is shown that slip at the wall in unsteady flows is governed by the gradient of shear rate and shear rate of the fluid. Non-dimensionalizing the model gives a universal curve which can be used to find the slip boundary condition at the fluid-solid interface based on the information of shear rate and gradient of shear rate of the fluid. A governing non-dimensional number, defined as the ratio of phase speed to speed of sound, is identified to help in explaining the mechanism responsible for the transition of slip boundary condition from finite to a perfect slip and determining when this would occur. Phase lag in fluid velocity relative to wall is observed. The lag increases with decreasing time period of wall oscillation and increasing wall hydrophobicity. The phenomenon of hysteresis is seen when looking into the variation of slip velocity as a function of wall velocity and slip velocity as a function of fluid shear rate. The cause for hysteresis is attributed to the unsteady inertial forces of the fluid.
\end{abstract}


\maketitle


\section{\label{sec:level 1}Introduction}
The no-slip boundary condition at the interface of a fluid and a solid wall  has been the subject of many investigation for more than a century~\cite{GoldsteinS:38a,GoldsteinS:69a,VinogradovaOI:99a}. Navier~\cite{NavierCLMH:1823a} was the first to introduce the linear boundary condition, which was later also proposed by Maxwell~\cite{MaxwellJC:90a}, and it remains the standard characterization of slip even today. Slip at the boundary although prevalent, is negligibly small in most continuum and macro-scale applications. Hence, the no-slip boundary condition is widely accepted and has been shown to give accurate results in such applications. However, in many micro- and/or nano-scale applications the first breakdown of continuum assumption often occurs at a solid boundary in the form of velocity slip. 

As transport in an ever smaller scales are considered, surface forces and effects start playing a more profound role on fluid transport than the bulk forces. Results from various computer simulations~\cite{KoplikJ:88a,KoplikJ:89a,ThompsonP:90a,TroianSM:97a,BarratJ:99a,LandmanU:00a,RobbinsMO:01a,TroianSM:04a,PriezjevNV:07a,PriezjevNV:07b,PriezjevNV:09a,PriezjevNV:10a}, which has been backed up by a number of laboratory experiments~\cite{IsraelachviliJN:96a,ChuraevNV:99a,CraigV:01a,ZhuY:01a,ButtHJ:02a,MazuyerD:02a,MeinhartCD:02a,LeeHJ:03a} show the presence of slip in fluids at the boundary. The advent of molecular dynamic simulations proved to be a considerable aid in understanding slip, as performing experiments at such scales is difficult. Most MD simulations have been focused on steady flows. MD simulations of a shear-driven steady flow by Thompson and Robbins ~\cite{ThompsonP:90a} showed dependence of slip on wall-fluid properties such as density of the wall relative to fluid, the strength of liquid-solid coupling, and the thermal roughness of the interface. They also observed layering of fluid normal to solid walls~\cite{ThompsonP:90a}. Thompson and Troian ~\cite{TroianSM:97a} performed MD simulation of the steady Couette flow. They observed that at small shear rates the boundary condition is consistent with the Navier model. However, as the shear rate is increased the Navier condition breaks down and the slip length increases rapidly with shear rate. They discovered that as the wall velocity is increased the slip at the wall is nonlinearly increased to infinity at a critical shear rate at which point the wall is no longer able to impart any further momentum to the fluid. They also went on to find a universal curve that gives the slip length for a specific shear rate irrespective of wal-fluid properties~\cite{TroianSM:97a}.

Boundary slip has been the subject of less investigation in unsteady flows. A few exceptions are in  unsteady gas flows~\cite{HillJM:09a,VafaiK:04a} and analytic solutions for continuum scale problems~\cite{BeskokA:04a,ZhangYH:08a,WangCY:11a,HillJM:07a}. However, to the authors knowledge, for liquids at micro-scales the research has been limited to steady flows. To this end, channel flows with oscillatory wall movement, the so-called Stokes second problem, appears as a natural extension of steady Couette flows investigated by Thompson and Troian. Such a flow can be encountered in several microsystem applications, such as microaccelerometers, inertial sensors, and resonant filters~\cite{KarniadakisG:05a}. 

In this paper the effects of unsteady flow on slip in simple liquids at the solid interface are presented. The numerical experiments conducted indicate that slip velocity is also dependent on fluid acceleration, in addition to the shear rate of the fluid. Previous studies on unsteady flows in microchannels by Khaled and Vafai~\cite{VafaiK:04a}, and Matthews and Hill~\cite{HillJM:09a} have used the Navier slip model, while Bahukudumbi et al.~\cite{BeskokA:03a} used the Maxwell slip model for their analysis. Both Navier and Maxwell's model suggest the dependence of slip only on shear rate. To confirm our hypothesis, Maxwell's theory for slip in rarefied gases~\cite{MaxwellJC:90a} for a  steady flow is extended to an unsteady case. Although the original derivation is inspired based on some characteristics of gases, we will verify here that an analogous formulation is also valid for liquid flows. The unsteady slip model developed shows the existence of an additional acceleration term. Using the momentum equation the acceleration term can be rewritten as gradient of shear rate. Hence showing that slip is dependent on shear rate and its gradient. Scaling the model by wall velocity and characteristic length for Stokes second problem, collapses the data to give a universal boundary condition for unsteady and steady flows. It is seen that at the limiting case when unsteady flow tends to steady flow the model reduces back to Navier and Maxwell's model. We also introduce a non-dimensional number that helps in explaining the transition of slip boundary condition from finite slip to perfect slip and determining when this would arise. Furthermore, the occurrence of hysteresis in unsteady flows is shown. Hysteresis is observed when comparing slip velocity with shear rate and the fluid velocity with wall velocity.


Details of numerical experiments and code validation are specified in sections~\ref{sec:numerical_setup} and~\ref{sec:validation}. In section~\ref{sec:unsteady_model} unsteady slip model is derived. Non-dimensionalization of the model is discussed in section~\ref{sec:nondimensionalization}. Results and conclusion are
made in section~\ref{sec:results} and~\ref{sec:conclusion} respectively.
\section{Numerical Setup}
\label{sec:numerical_setup}

\begin{figure}[h!]
 \includegraphics[width=0.25\textwidth]{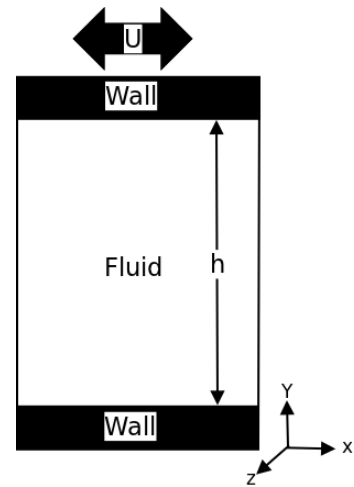}
 \caption{Schematic of the problem geometry, where $h$ is the height of the fluid channel and $U$ is the wall velocity.}
 \label{f:Schematic}
\end{figure}

The molecular dynamic simulations presented in this paper are performed using the LAMMPS package~\cite{PlimptonS:95a}. The problem geometry used is similar
to that of Stokes second problem which is achieved by selecting the height of the fluid channel to be greater than the Stokes penetration depth as shown in Fig.~\ref{f:Schematic}. The 
penetration depth as determined from the analytical solution of the Stokes problem, is given by $\delta=6.51\sqrt{{\nu}/{\omega}}$, where $\nu$ is the kinematic viscosity of the fluid and $\omega$ is the frequency of wall oscillation.
The channel dimensions and number of wall and fluid atoms for different wall densities
and various time periods are presented in Table~\ref{tab:myfirsttable}. 
\begin{table}[h!]
\begin{tabular}{l*{6}{c}r}
  \hline
  \hline
  &$T/\tau$~  & ~$\rho_w/\rho$~ & ~$x/\sigma$~ & ~$y/\sigma$~ & ~$z/\sigma$~  & ~$N_f$~ & ~$N_w$\\
  \hline
  &$\le120$~         & ~$1$~ & ~$11.95$~ & ~$50$~ & ~$1204.14$~ &  ~$595,000$~ &  ~$23,100$\\
  &$\le120$~         & ~$4$~ & ~$12.14$~ & ~$50$~ & ~$1204.14$~ & ~$595,000$~ & ~$58,656$\\
  &$\le400$~         & ~$1$~ & ~$11.95$~ & ~$100$~ & ~$1204.14$~ &  ~$1190,000$~ &  ~$23,100$\\
  &$\le400$~         & ~$4$~ & ~$12.14$~ & ~$100$~ & ~$1204.14$~ & ~$1190,000$~ & ~$58,656$\\
  &$800$~         & ~$1$~ & ~$11.95$~ & ~$150$~ & ~$1204.14$~ &  ~$1785,000$~ &  ~$23,100$\\
  &$800$~         & ~$4$~ & ~$12.14$~ & ~$150$~ & ~$1204.14$~ & ~$1785,000$~ & ~$58,656$\\
\hline
\hline
\end{tabular}
\caption{\label{tab:myfirsttable} Dimensions of the fluid domain and the number of fluid and wall atoms are enlisted for different values of time period. Time period $T$, wall number density $\rho_w$, length $x$, height $y$, and width $z$ of the fluid channel, number of fluid $N_f$ and wall $N_w$ atoms are the variables mentioned. Here $\tau$ is characteristic time of the Lennard-Jones potential and $\rho$ is the fluid number density. The minor changes in the dimensions with change in wall density are done to make the problem geometry symmetric and periodic.}
\end{table}
The fluid's initial state is modeled as a face centered cubic (fcc) structure with the $x$ direction
of the channel being aligned along the $[11\bar{2}]$ orientation of the face-centered cubic lattice.
The wall is comprised of two layers of atoms oriented along the $(111)$ plane of fcc lattice. The wall
atoms are fixed to their lattice sites. The bottom wall is kept stationary while the top wall
is subjected to an oscillatory motion, defined by
\begin{equation}
 x=Asin(\omega t),
\end{equation} 
where $A$ is the amplitude of wall oscillation. Periodic boundary
conditions are imposed along the $x$ and $z$ directions. 

The pairwise interaction of atoms separated by a distance $r$ is modeled by the Lennard Jones potential
\begin{equation}
V^{LJ}=4\epsilon\left[\left(\frac{\sigma}{r}\right)^{12}-\left((\frac{\sigma}{r}\right)^{6}\right],
\end{equation}
where $\epsilon$ and $\sigma$  are the characteristic energy and length scales.
The cutoff radius, $r_c$ is $2.2 \sigma$ where the potential is zero for $r>r_c$.

The fluid is maintained in its equilibrium state having a number density $\rho=0.81\sigma^{-3}$ and temperature $T=1.1k_B/\epsilon$.
The temperature is regulated by a thermostat which simulates the
transfer of heat from the system to an external reservoir. A Langevin thermostat with a damping coefficient of $\Gamma=1.0\tau^{-1}$, where $\tau=\sqrt{m\sigma^{2}/\epsilon}$, is used here. The damping term is only applied
to the $z$ direction to avoid biasing the flow. The equation of motion of the fluid atom of mass $m$
along the $z$ component is therefore given as follows
\begin{equation}
 m\ddot{z_i}=\sum_{j\neq i}\frac{\partial V{ij}}{\partial z{i}}-m\Gamma \dot{z_i} + \eta_i.
\end{equation}
Here $\sum_{j\neq i}$ denotes sum over all interactions with i and $\eta_i$ is a Gaussian distributed random force. The value of dynamic viscoisty used for the calculations is $\mu=0.2 \epsilon\tau\sigma^{-3}$. The equations of motion were integrated using the Verlet algorithm~\cite{VerletL:67a,TildesleyDJ:87a} with a time step $\tau_c=0.002\tau$.

The simulation is initially run for a time of $\sim600\tau$ allowing the flow to equilibrate. After which, an ensemble average of required variables are taken in addition to spatial averaging. The spatial averaging is done along the length and width of the channel, with a bin height
of $0.25 \sigma$. 
\section{Validation of simulation}
\label{sec:validation}
\begin{figure}[h!]
  \includegraphics[width=0.43\textwidth]{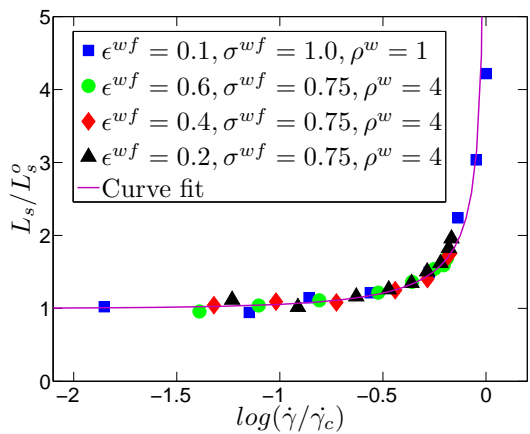}
  \caption{Universal curve describing the flow boundary condition. $L_s$ and $\dot{\gamma}$ is scaled by $L_s^0$ and $\dot{\gamma_c}$ respectively. The curve fit is given by $L_s=L_S^0(1-\dot{\gamma}/\dot{\gamma_c})^{-1/2}$ in agreement by Thompson and Troian's results~\cite{TroianSM:97a}.}
  \label{f:Char_curve}
\end{figure}
Before going ahead with our present experiment we validated our code by reproducing Thompson and Troian's results~\cite{TroianSM:97a}.
The problem consisted of a Couette flow geometry. Simulation was run for different cases having varying
wall-fluid properties. For each case, slip length was computed for increasing shear rate of fluid which
in turn is governed by the wall velocity. The slip length is calculated from the linear Navier boundary condition,
 $\Delta V=L_s\dot{\gamma}$, which can be simplified for Couette flow to $(U/\dot{\gamma}-h)/2$~\cite{TroianSM:97a}, where $\Delta V$ is the slip velocity,~$L_s$ is the slip length and~$\dot{\gamma}$ is the fluid shear rate. The shear rate is computed as the slope of the velocity profile. The reproduced results of Thompson and Troian~\cite{TroianSM:97a} is shown in Fig.~\ref{f:Char_curve}. Considerable agreement was found with their results. The non linear dependence of slip length with shear rate is illustrated. Also we were able to duplicate the universal curve~\cite{TroianSM:97a} which is a plot of slip length normalized by its limiting value,~$L_S^o$, versus the shear rate which is normalized by the critical shear rate,~$\dot{\gamma_c}$. This primarily ends up collapsing the data onto one curve given by $L_s=L_s^o(1-\dot{\gamma}/{\gamma_c})^{-1/2}$. The universal curve shows that for a given shear rate the non-dimensionalized slip length is independent of the fluid wall properties of the problem being considered. 
\section{Unsteady slip model}
\label{sec:unsteady_model}
The two widely used model namely Navier's and Maxwell's slip model are in essence the same. Both assume that the slip velocity at the fluid-solid interface is proportional to the shear rate of the fluid. Even though Maxwell's model ~\cite{MaxwellJC:90a,LoebLB:34a,GombosiTI:94a} was established for rarefied gases we illustrate here that an analogous formulation can be made for slip in liquids.

\begin{figure}[h]
\includegraphics[width=0.43\textwidth]{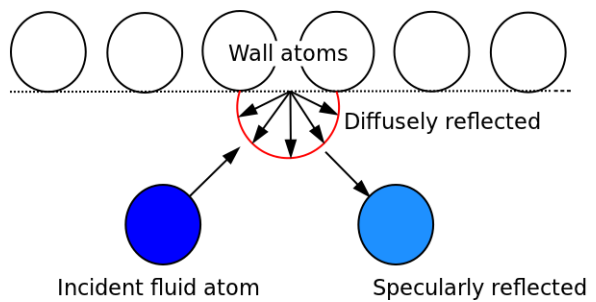}
 \caption{Schematic of diffusively and specularly reflected fluid atom.}
 \label{f:Reflection}
\end{figure}

Maxwell's theory states that the reflection of fluid atoms after colliding with wall atoms can be categorized into two types.
The two types being specular reflection and diffusive reflection as shown in Fig.~\ref{f:Reflection}. 

\textit{Diffusive reflection}~$\left(U_w^{f1}\right)$:
In this case the incident fluid atoms exhibit no-slip and the velocity of the atom attained after collision is the same as that of the velocity of the wall
during collision. This can be imagined as fluid atoms being adsorbed by the wall and then put back into
the fluid in a random direction. Hence, the aggregate velocity of the atom after collision being equal 
to the velocity of wall atom at the instant of collision (relative velocity with wall being zero)
\begin{equation}
U_w^{f1}(t_c^+)=U^w(t_c).
\label{eq:nd_1}
\end{equation}
Here $t_c$ is the instantaneous time of collision and $U^w$ is the wall velocity.

\textit{Specular reflection}~$\left(U_w^{f2}\right)$:
The atoms undergo perfect slip and the velocity after collision is the same as that before collision
\begin{equation}
U_w^{f2}(t_c^+)=U_w^f(t_c^-),
\label{eq:nd_2}
\end{equation}
where $U_w^f$ is the fluid velocity at the wall.
Furthermore, the velocity of the atom before collision with wall is deemed to have been obtained by collision with fluid atoms located at a distance equal to $\lambda$ from the wall. This being the location of the first fluid layer and $\tau_c$ is the time it takes for the fluid atom to travel this distance. Here we make an assumption that during the motion of the fluid atom in reference it is subjected to a net zero force during its journey from the first fluid layer to the wall. Thereby, it has a constant velocity
\begin{equation}
U_w^f(t_c^-)=U_{\lambda}^f(t_c-\tau_c).
\end{equation}
Performing a Taylor series expansion in space at the wall the velocity at a distance $\lambda$ from it can be written as
\begin{equation}
  U_{\lambda}^f(t_c-\tau_c)=U_w^f\left({t_c}-\tau_c\right)-\lambda\frac{dU^f\left(t_c-\tau_c\right)}{dy} \Bigr\rvert_{w}.
\end{equation}
Discretizing in time and ignoring all second and higher order terms we obtain
\begin{equation}
  U_w^{f}(t_c^-)=U_{\lambda}^f(t_c-\tau_c)=U_w^f\left(t_c\right)-\tau_c\frac{dU_w^f \left(t_c \right)}{dt}  - \lambda\frac{dU^f\left(t_c\right)}{dy} \Bigr\rvert_{w}.
\label{eq:nd_5}
\end{equation}

Now, the average velocity after collision $U_w^f(t_c^+)$ is given by,
\begin{equation}
  {U_w^f(t_c^+)}=\sigma_d U_w^{f1}(t_c^+) + \left(1-\sigma_d\right)U_w^{f2}(t_c^+),
\end{equation}
where $\sigma_d$ is called the tangential momentum accommodation coefficient (TMAC), which gives the fraction of atoms undergoing diffusive reflection. Substituting from equations (\ref{eq:nd_1}) and (\ref{eq:nd_2}) in the above equation we obtain
\begin{equation}
  {U_w^f(t_c^+)}=\sigma_d U^w\left({t_c}\right) + \left(1-\sigma_d\right)U_w^{f}(t_c^-).
\label{eq:nd_7}
\end{equation}

The net instantaneous fluid velocity at the wall is given as the mean of the  velocity before and after collision
\begin{equation}
  U_w^f(t_c)=\frac{U_w^f(t_c^+)+U_w^f(t_c^-)}{2}.
\label{eq:nd_8}
\end{equation}
By substituting $U_w^f(t_c^+)$ from (\ref{eq:nd_7}), one obtains
\begin{equation}
  2U_w^f(t_c)=\sigma_d U^w\left({t_c}\right) + \left(2-\sigma_d\right)U_w^{f}(t_c^-).
\label{eq:nd_9}
\end{equation}
Now $U_w^{f}(t_c^-)$ could be replaced from equation (\ref{eq:nd_5}) to give
\begin{align}
2 U_w^f\left(t_c\right)&=\sigma_d U^w\left({t_c}\right)+2U_w^f\left(t_c\right)-\sigma_d U_w^f\left(t_c\right)+\notag\\ &\left(2-\sigma_d\right)
\left[-\tau_c\frac{dU_w^f \left(t_c \right)}{dt}  - \lambda\frac{dU^f\left(t_c\right)}{dy} \Bigr\rvert_{w}\right].\notag\\
\label{eq:nd_10}
\end{align}
Rearranging this equation results in
\begin{equation}
U^w\left({t_c}\right)- U_w^f\left(t_c\right)=\frac{\left(2-\sigma_d\right)}{\sigma_d}\left[\tau_c\frac{dU_w^f 
\left(t_c \right)}{dt}  + \lambda\frac{dU^f\left(t_c\right)}{dy} \Bigr\rvert_{w}\right].
\label{eq:nd_11}
\end{equation}

Slip velocity is given as, 
\begin{equation}
U_s=U^w-U_w^f.
\label{eq:nd_12}
\end{equation}
Using the definition of slip velocity equation (\ref{eq:nd_11}) can be written as
\begin{equation}
 U_s\left({t_c}\right)=\frac{\left(2-\sigma_d\right)}{\sigma_d}\left[\tau_c\frac{dU_w^f \left(t_c \right)}{dt}  + 
\lambda\frac{dU^f\left(t_c\right)}{dy} \Bigr\rvert_{w}\right].
\label{eq:nd_13}
\end{equation}
Finally writing $dU/dy$ in terms of shear rate, $\dot{\gamma}$ we get
\begin{equation}
 U_s\left({t_c}\right)=\frac{\left(2-\sigma_d\right)}{\sigma_d}\left[\tau_c\frac{dU_w^f \left(t_c \right)}{dt}  + 
\lambda\dot{\gamma} \Bigr\rvert_{w}\right].
\label{eq:nd_14}
\end{equation}
Hence, we see that slip velocity has an additional dependence on the fluid acceleration at the wall in case of unsteady flows. 

The simplified Navier Stokes equation is used in order to rewrite the acceleration term in the above expression in terms of shear rate. Writing down the 2-D Momentum equation in $x$, with no external force or pressure gradient
\begin{equation}
\frac {\partial u}{\partial t} + u\frac {\partial u}{\partial x}+ v\frac {\partial u}{\partial y} =
+ \nu \left(\frac {\partial^2 u}{{\partial x}^2} + \frac {\partial^2 u}{{\partial y}^2}\right).
\end{equation}
In addition since the velocity of fluid in the $x$ direction is uniform, the rate of change of velocity along $x$ direction goes to zero. Hence the Navier Stokes equation reduces to,
\begin{equation}
 \frac {\partial u}{\partial t} =\nu \left( \frac {\partial^2 u}{{\partial y}^2}\right).
\end{equation}
\noindent Now considering that,~$\dot\gamma=\frac {\partial U_w^f}{\partial y}$ one can rewrite the above equation as,
\begin{equation}
 \frac {\partial U_w^f}{\partial t} =\nu \frac {\partial }{\partial y}\left(\frac {\partial U_w^f}{\partial y}\right)=\nu \left( \frac {\partial {\dot\gamma}}{{\partial y}}\right).
\end{equation}
\noindent Substituting this in the equation for slip velocity, equation (\ref{eq:nd_14}), results in
\begin{equation}
 U_s\left({t_c}\right)=\frac{\left(2-\sigma_d\right)}{\sigma_d}\left[\tau_c\nu \left( \frac {\partial {\dot\gamma}}{{\partial y}}\right)  + 
\lambda\dot{\gamma}\right]_{w}.
\end{equation}

For steady flow, slip velocity is seen to be proportional to shear rate of the fluid as hypothesized by Navier and Maxwell. But here, we show that for an unsteady problem we have an additional contribution from the gradient of shear rate. Hence it is seen that slip velocity for an unsteady case is proportional to the linear sum of shear rate and gradient of shear rate of the fluid at the wall.

In the limitng case when the flow approaches steady state and the acceleration goes to zero the unsteady slip model reduces back to that given by Maxwell,
\begin{equation}
 U_s=\left(\frac{2-\sigma_d}{\sigma_d}\right)\lambda\dot{\gamma}
\end{equation}
%
\noindent Comparing this to Navier's slip bondary condition
\begin{equation}
 U_s=L_s\left.\dot{\gamma}\right|_{w},
\end{equation}
one can write the slip length, $L_s=\frac{\left(2-\sigma_d\right)}{\sigma_d}\lambda.$
A $\sigma_d$ value of one can be compared to a no-slip boundary condition in liquids where there is no relative velocity between the wall and fluid. While $\sigma_d$ equal to zero corresponds to a boundary condition exhibiting perfect slip.
%
%
%
\section{Non-dimensionalizing the unsteady slip model}
\label{sec:nondimensionalization}
%
Non-dimensionalization of the slip model is performed with an attempt to simplify and parametrize the equation. It is also aimed at
identifying a non-dimensional parameter that could characterize unsteady slip. The unsteady slip model established in previous section is scaled by the wall velocity, $U_w$, and the length scale associated with Stokes second problem, $\sqrt{{2\nu}/{\omega}}$. Non-dimesnioalizing using this velocity and length scales, one obtains
\begin{equation}
 U_s^*=\left(\frac{2-\sigma_d}{\sigma_d}\right)
\left[(\tau_c\nu)\frac{1}{\frac{2\nu}{\omega}} \left( \frac {\partial^2 U_f^*}{\partial y^{* 2}}\right)  + 
\frac{\lambda}{\sqrt{\frac{2\nu}{\omega}}}\frac{\partial U_f^*}{\partial y^*} \right]_{w},
\end{equation}
where $(.)^*$ represents non-dimensional quantities. Notice that,
$\tau_c=\frac{\lambda}{\bar{c}}$, where $\bar{c}$ is the speed of sound in the fluid at a given temperature. Therefore
\begin{equation}
 U_s^*=\left(\frac{2-\sigma_d}{\sigma_d}\right)\lambda^*
\left[C_{nd}  \frac {\partial^2 U_f^*}{\partial y^{* 2}}  + 
\frac{\partial U_f^*}{\partial y^*} \right]_{w}.
\end{equation}
This could be also written as,
\begin{equation}
  U_s^*=L_s^*
  \left[C_{nd}  \frac {\partial^2 U_f^*}{\partial y^{* 2}}  + 
  \frac{\partial U_f^*}{\partial y^*} \right]_{w}.
\label{eq:nond_1}
\end{equation}
where $L_s^*=\left(\frac{2-\sigma_d}{\sigma_d}\right)\lambda^*$ and $C_{nd}=\frac{\sqrt{2\nu\omega}}{2\bar{c}}$ is the non-dimensional parameter that controls the unsteady term based on the type of flow. One can see that for a limiting case of $\omega=0$, which corresponds to a steady flow problem, the equation reduces back to Maxwell's slip equation.
%
\section{Results}
\label{sec:results}
The primary information that is extracted from the simulations is the fluid velocity. The following steps are taken before the data from the simulations is used for any analysis. First, a reference plane is defined at a distance of $0.5 \sigma^{wf}$ away from the wall lattice site. This is the location at which the fluid variables at the wall is computed. Secondly in order to obtain a well resolved velocity profile the Levenberg-Marquardt method~\cite{FletcherR:71a,BaldaM:07a} is used to fit the analytical solution of the Stokes second problem to the data. This is done so as to limit the noise resulting from taking the derivatives of the velocity profile. Different cases of wall-fluid properties considered are enlisted in Table~\ref{tab:mysecondtable}.
\begin{table}[h!]
\begin{tabular}{l*{4}{c}r}
  \hline
  \hline
  & Case & ~$\epsilon^{wf}/\epsilon$~ & ~$\sigma^{wf}/\sigma$~ & ~$\rho_w/\rho$~ \\
  \hline
  &1~ &~$0.6$~         & ~$1.0$~ & ~$1$~ \\
  &2~ &~$0.1$~         & ~$1.0$~ & ~$1$~ \\
  &3~ &~$0.4$~         & ~$0.75$~ & ~$4$~ \\
  &4~ &~$0.2$~         & ~$0.75$~ & ~$4$~ \\
\hline
\hline
\end{tabular}
\caption{\label{tab:mysecondtable} Four different cases with varying wall-fluid properties were considered.~$\epsilon^{wf}$ and $\sigma^{wf}$ are the Lennard-Jones parameters for fluid-wall interaction.}
\end{table}
 
Thompson and Troian~\cite{TroianSM:97a} observed that slip length for fluid shear rates much lower than the critical shear rate is constant. But for shear rates in the vicinity of the critical value it becomes highly nonlinear~\cite{TroianSM:97a}. As we are here considering an unsteady problem the various fluid and wall variables are time varying, and so is slip length if the shear rate is in the vicinity of its critical value. Hence, the slip length corresponding to each run is calculated by fitting the data using the Lavenberg-Marquardt method~\cite{FletcherR:71a,BaldaM:07a} and the slip model given in equation~\ref{eq:nond_1}. 
\subsection{Verification of slip model}
\begin{figure}[h!]
\subfigure[~]{\label{fig:Ls_T}\includegraphics[width=0.43\textwidth]{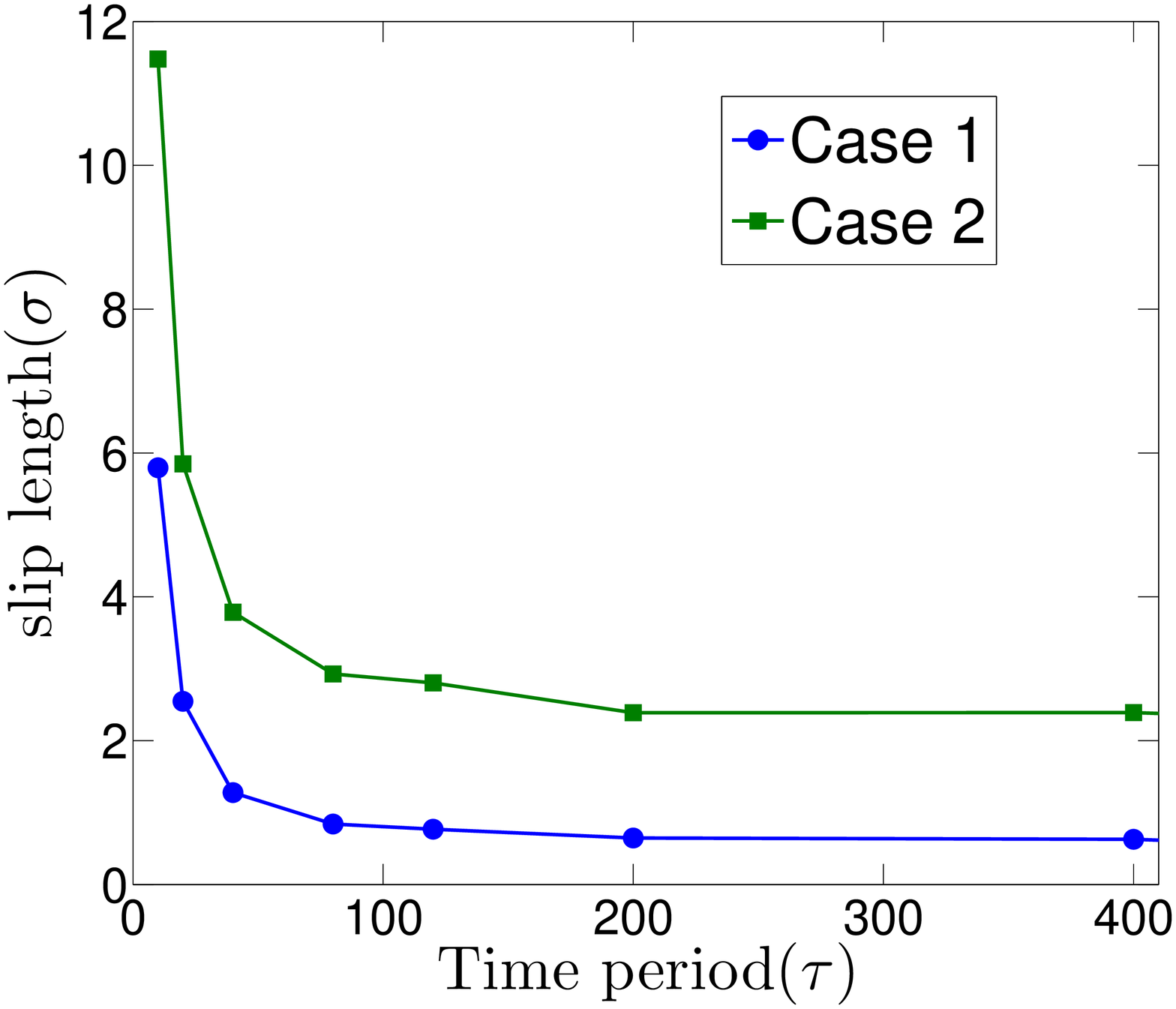}}
\subfigure[~]{\label{fig:Ls_coll}\includegraphics[width=0.43\textwidth]{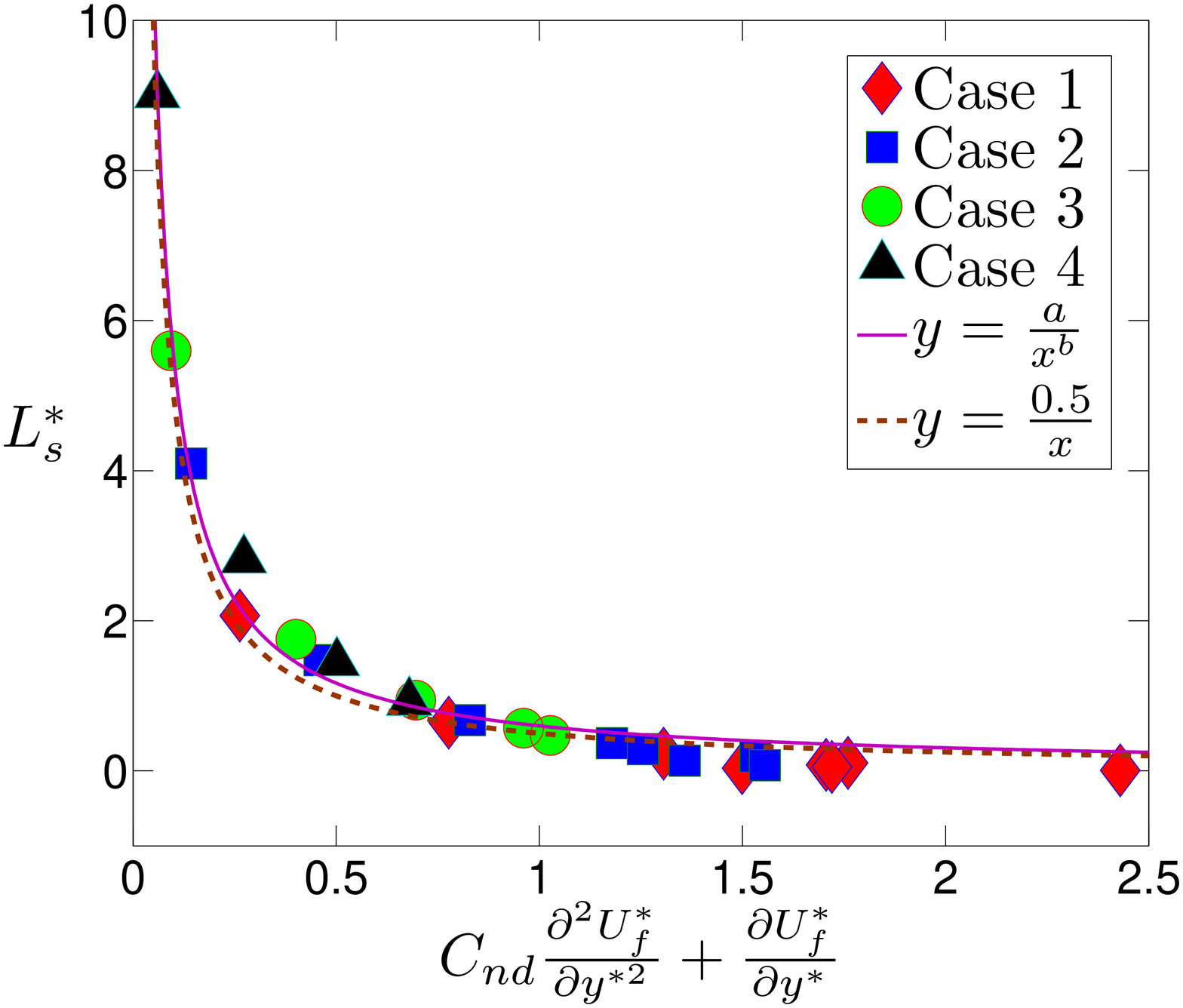}}

 \caption{(a) Slip length versus time period. As the unsteady problem tends to a steady problem, steady state value of slip length is recovered. (b) Non-dimensional slip length versus sum of the non-dimensional shear rate and the product of non-dimensional number $C_{nd}$ and gradient of shear rate. Here $y=L_s^*$ and $x=C_{nd}\frac{\partial^2U_f^*}{\partial y^{*2}}+\frac{\partial U_f^*}{\partial y^*}$. The curve fitting parameters are $a=0.59$ and $b=0.96$. It is seen that all the data for various cases considered collapse onto a single curve given by the fit.}
 \label{fig:Ls}
\end{figure}
Several numerical simulations were conducted in order to verify the unsteady slip model established in the previous section. The variation of slip length with time period, shown in Fig.~\ref{fig:Ls}(a), gives us considerable insight into the unsteady problem. It is seen that the slip length approaches infinity as the wall oscillation frequency tends to infinity. This implies that the wall is oscillating at such a high frequency that no information can be passed on by the wall atoms to the underlying fluid atoms, thereby giving a perfect slip. As we increase the time period we observe that the slip length decreases and approaches an asymptotic finite value which corresponds to those shown by Thompson and Troian in the steady state Couette flow problem~\cite{TroianSM:97a}.

In Fig.~\ref{fig:Ls}(b) the non-dimensionalized slip length, $L_s^*$ also mentioned as $y$, is plotted against $x$ which is given as  $\left[C_{nd}\frac {\partial^2 U_f^*}{\partial y^{* 2}}  + \frac{\partial U_f^*}{\partial y^*} \right]_{w}$ where $x$ is a sum of the parameters that influence unsteady slip. The data for different cases of wall-fluid properties and different wall velocities collapse onto a single curve. This curve is well described by the form $y={a}/{x^{b}}$ where the dashed line represents the values $a=0.5$ and $b=1.0$. The exact fitting coefficients to our experimental data were calculated to be $a=0.59$ and $b=0.96$, given by solid line which closely matches the dashed line. Hence, it is seen here that similar to Thompson and Troian proper scaling leads to a universal curve which holds for unsteady and steady flows. With the help of this curve one can find the slip length for a given value of shear rate and its gradient.
\subsection{Relevance of Non-dimensional number}
In the non-dimensional number $C_{nd}$, $\sqrt{2\nu\omega}$ is the phase velocity of the propagating velocity profile in $y$-direction whereas $\bar{c}$ is the speed of sound through the medium. The ratio can be physically seen as the ratio of propagation of momentum through the medium to the speed of propagation of sound through the medium.  The maximum speed at which momentum can be transferred through a medium is at the speed of sound. Hence, when the phase speed is of the order of the speed of sound for the medium, the fluid will exhibit perfect slip as the wall would not be able to transfer momentum on to the fluid. The number can be used to determine when the boundary condition would change to perfect slip from a finite or no-slip condition.

\subsection{Effects of acceleration on slip}
\begin{figure}[h]
 \includegraphics[width=0.43\textwidth]{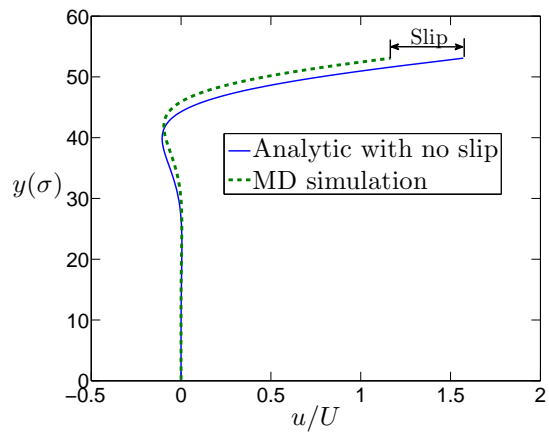}
 \caption{Velocity profile obtained from simulation with Stokes analytical solution for no-slip boundary condition is compared. The profile is for case 1 of wall-fluid properties and for an instant of time equal to the time period of wall oscillation. A distinct presence of slip at the wall is seen.}
 \label{f:Vel_profile}
\end{figure}

In order to analyze the affect of acceleration on slip a wall-fluid property is chosen which is shown to exhibit no-slip at the fluid-solid interface for steady flow by Thompson and Troian~\cite{TroianSM:97a}. This corresponds to case 1 in Table~\ref{tab:mysecondtable}. Wall oscillation having an amplitude of $10\sigma$ and time period of $40\tau$ is chosen such that the shear rate is considerably below its critical value. In Fig.~\ref{f:Vel_profile}, comparison of the velocity profile obtained from simulation is made with Stokes analytical solution with no slip boundary condition. This done for an instantaneous time $t=40\tau$ for which distinct slip at the wall is observed.
\subsubsection{Phase lag of fluid velocity due to wall acceleration}
\begin{figure}[h]
 \includegraphics[width=0.43\textwidth]{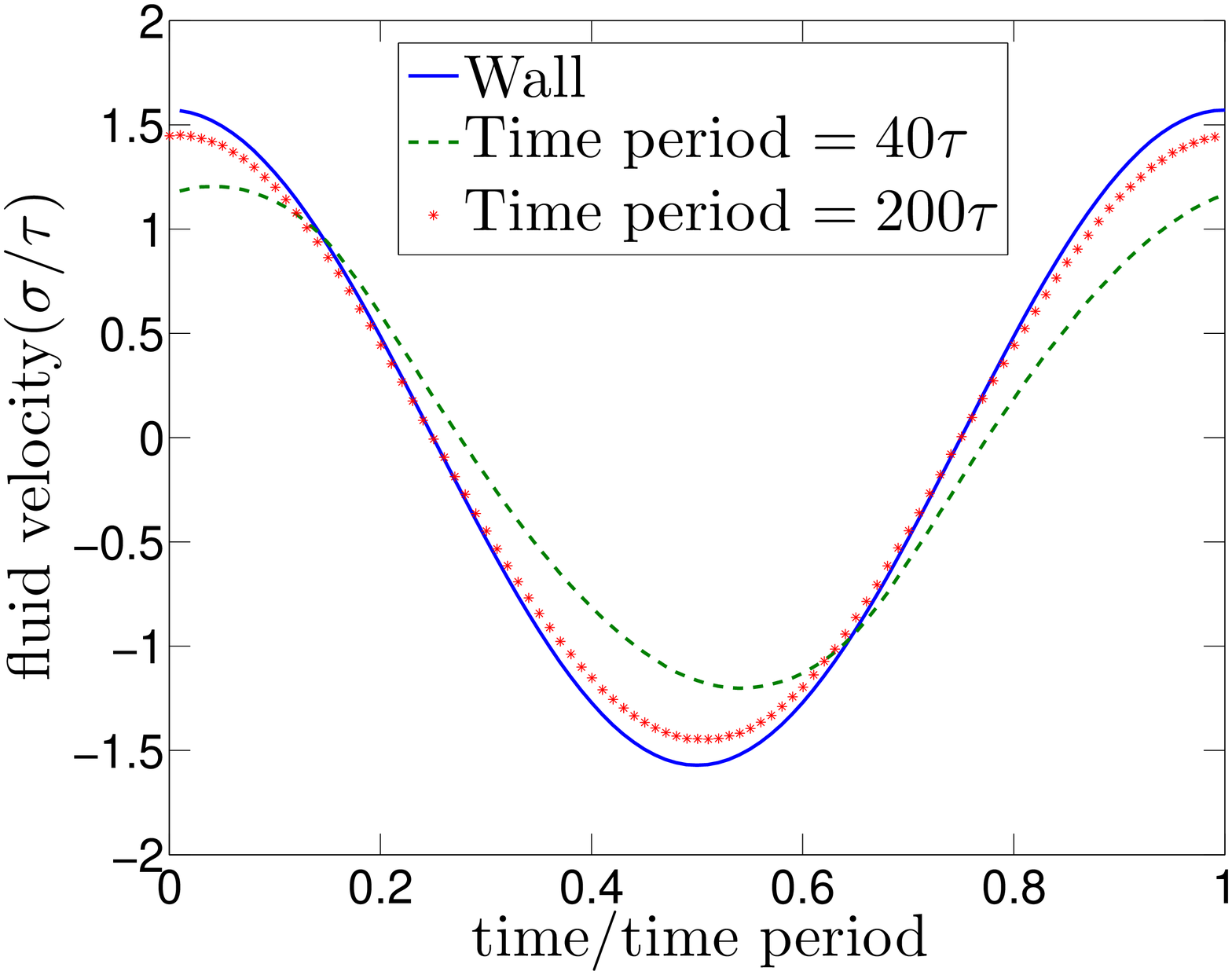}
 \caption{Comparison of wall velocity and fluid velocities at time periods $40\tau$ and $200\tau$ for a cycle of wall oscillation having wall-fluid properties corresponding to case 1. An amplitude of $10\sigma$ and $50\sigma$ corresponding to the time periods of $40\tau$ and $200\tau$ are chosen in order to achieve the same amplitude of wall velocity in both cases. An increase in wall-fluid phase lag and a decrease in the amplitude of fluid adjacent to the wall are observed by decreasing time period.}
 \label{f:Ph_lag_1}
\end{figure}

The presence of fluid slip at the wall indicates the existence of a lag of fluid velocity with respect to the wall. This is investigated by comparing fluid velocity for varying time periods of wall oscillation with the wall velocity as shown in Fig.~\ref{f:Ph_lag_1}. This is done to see the effects of acceleration on the fluid velocity. The comparison is made for case 1 of wall-fluid properties. The amplitude of wall oscillation is calculated such that the amplitude of wall velocity remains constant. It is seen that the phase lag reduces and the amplitude of fluid velocity increases with increasing time period relative to the wall. An increase in time period results in a decrease in fluid inertia. Hence, it can be said that as we approach a steady state, by increasing time period, we recover the no-slip boundary condition for case 1. Leading to the conclusion that inertia affects the slip of fluid at the wall. Thereby, confirming the general slip model derived in previous section.

\subsubsection{Phase lag of fluid velocity due to wall hydrophobicity}
\begin{figure}[h]
 \includegraphics[width=0.43\textwidth]{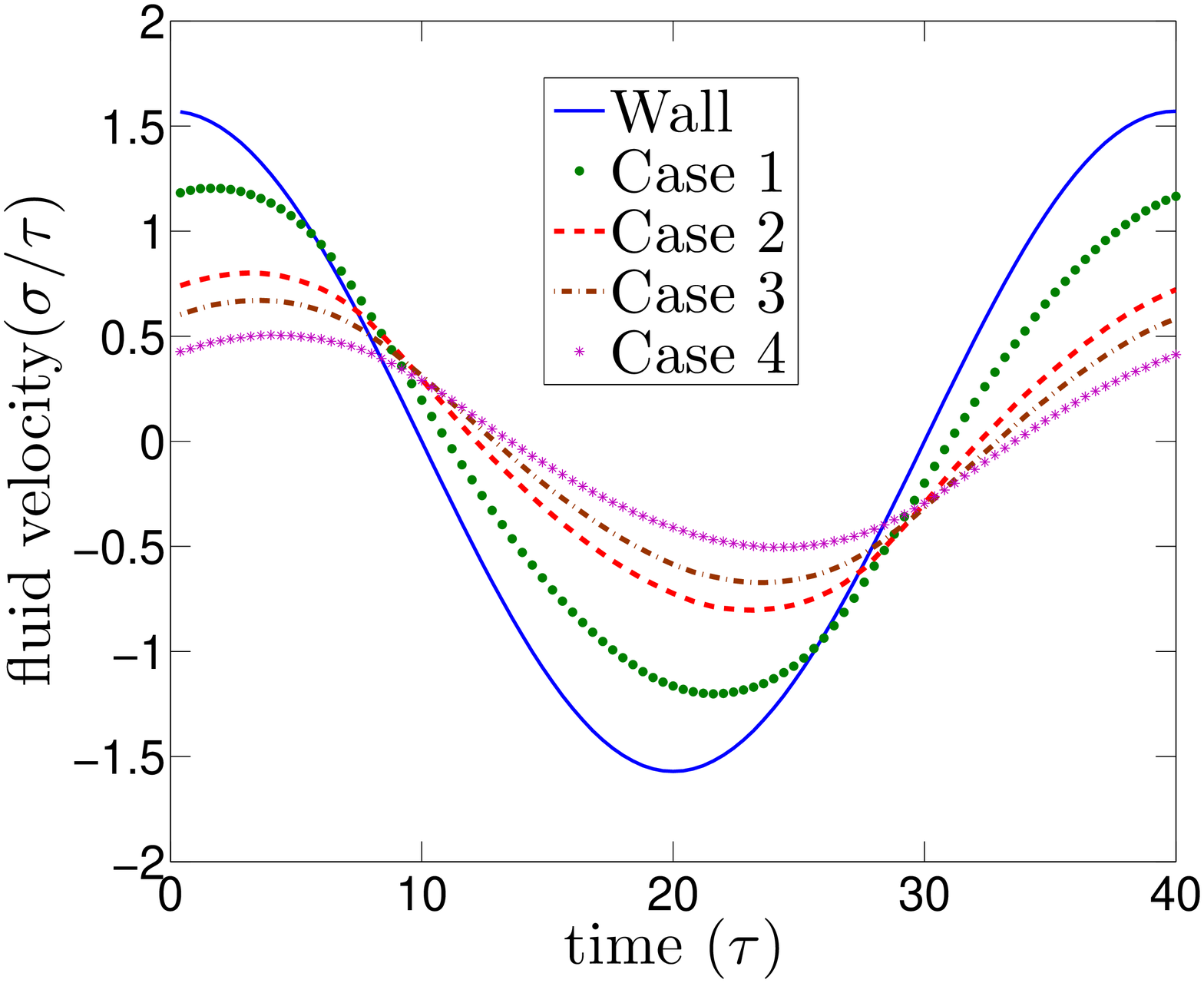}
 \caption{Velocities of wall and adjacent fluid in one wall oscillation cycle. The wall oscillation has an amplitude of $10\sigma$ and time period $40\tau$. An increase in wall-fluid phase lag and a decrease in the amplitude of fluid adjacent to the wall are observed by increasing the wall-fluid hydrophobicity.}
 \label{f:Ph_lag}
\end{figure}

The variation of fluid velocity with time for different cases mentioned in Table~\ref{tab:mysecondtable} is shown in Fig.~\ref{f:Ph_lag}. This is done for a fixed time period of $40\tau$ and fluid acceleration of $10\sigma$. The purpose being to see how the fluid velocity behaves with varying hydrophobicity. The fluid velocities lag with respect to the wall velocity. There is an increase in phase lag with increasing hydrophobicity. Also, a reduction in the amplitude of the fluid velocity is noticed. The phase lag and amplitude reduction are a result of the increase in slip. Tang et al ~\cite{ZhangYH:08a} also observed similar phase lag of fluid velocity for different Stokes numbers and TMAC in their Lattice Boltzmann simulation of oscillatory gas flows.

\subsection{An explanation for the occurrence of slip}
\begin{figure}[h]
 \includegraphics[width=0.43\textwidth]{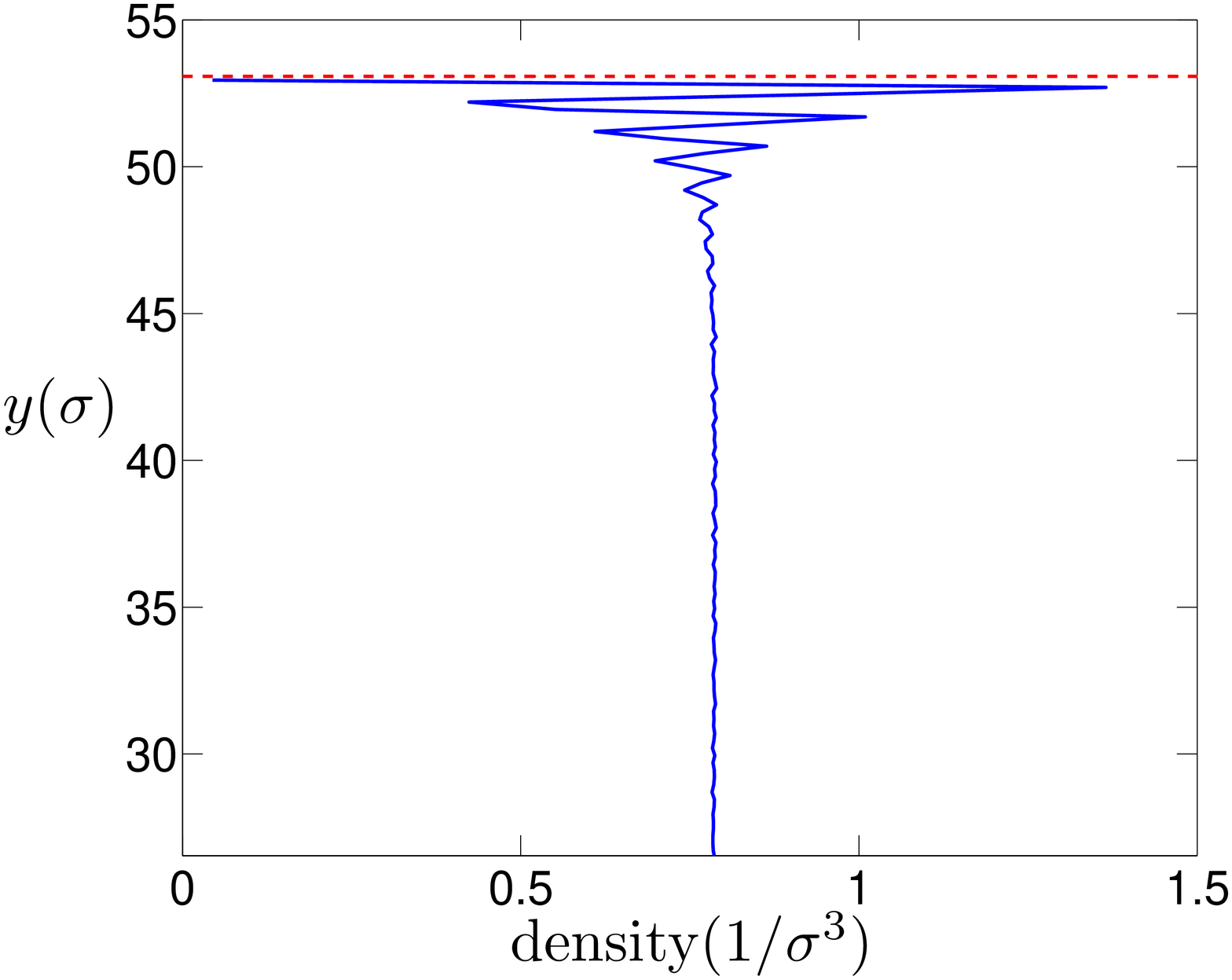}
 \caption{Variation of fluid density along the height of fluid channel. Layering of fluid is observed close to the wall. Red dashed line marks the reference plane $0.5\sigma^{wf}$ away from the wall lattice site. The plot corresponds to case 1 with an amplitude of $10\sigma$ and $40\tau$. As the layering is symmetric about $x-z$ plane only the top half is shown. The ordering is prevalent up to a distance of $5-6\sigma$ from the reference plane with decreasing amplitude beyond which the bulk density of the fluid is obtained.}
 \label{f:Dens_lay}
\end{figure}

Increase in hydrophobicity leads to an increase in slip at the wall, which has been observed by several researchers for the steady case~\cite{KoplikJ:88a,KoplikJ:89a,ThompsonP:90a,TroianSM:97a,BarratJ:99a,LandmanU:00a,RobbinsMO:01a,TroianSM:04a,PriezjevNV:07a,PriezjevNV:07b,PriezjevNV:09a,PriezjevNV:10a}. 
Strictly speaking, the no-slip boundary condition
is only valid if the flow adjacent to a solid surface is in thermodynamic equilibrium~\cite{HillJM:07a,HillJM:09a}. For fluid flow in small-scale
systems, the collision frequency is not high enough to ensure thermodynamic equilibrium, thus a certain degree of tangential
velocity slip must be allowed~\cite{HillJM:09a}. Also, these collisions should occur during a time interval smaller than that of the smallest time scale for flow changes.
 Harris and Rice calculated the relaxation time  for liquid argon at $90$ K as $0.5\tau$ while for gaseous argon at $300$ K to be of the order of $100\tau$~\cite{RiceSA:60a}. The relaxation time required for bulk liquid argon is small in comparison to the time scale of molecular collisions. However, as shown in Fig.~\ref{f:Dens_lay} fluid layering is observed close to the wall. This leads to the reduction of liquid density to the order of gas density adjacent to the wall. This reduces the number of fluid atoms interacting and undergoing momentum transfer with the wall thereby increasing the required relaxation time~\cite{ZhangYH:08a}. Hence, the fluid atoms adjacent to the wall do not have sufficient time to equilibrate and the transfer of momentum from the wall is only partial; therefore, resulting in slip.

\subsection{Boundary condition for an unsteady flow}
\begin{figure}[h]
 \includegraphics[width=0.43\textwidth]{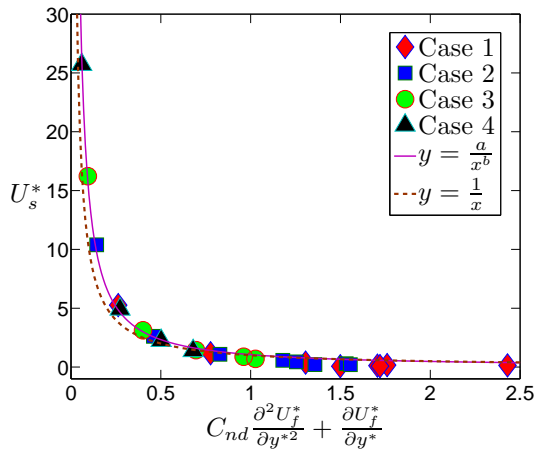}
 \caption{Data for different cases collapses onto a single curve defined by $y=\frac{a}{x^{b}}$, where $y=\tilde{U_s}$ and $x=C_{nd}  \frac {\partial^2 U_f^*}{\partial y^{* 2}}  + 
\frac{\partial U_f^*}{\partial y^*}$. The parameters for the curve fit are a=1.05 and b=1.14.}
 \label{f:v_collapse}
\end{figure}

The main aim of the studies pertaining to slip length is to find the exact boundary condition, which is slip velocity at the wall. We find that similar to slip length, slip velocity also collapses onto a single universal curve when non-dimensioanlized with the appropriate scaling parameters as shown in Fig.~\ref{f:v_collapse}.
The scaled slip velocity is defined as,
\begin{equation}
  \tilde{U_s}=U_s^*\left(\frac{U_w}{\dot{\gamma}\sqrt{\frac{2\nu}{\omega}}}\right),
\end{equation}
where $U_s^*$ is the slip velocity scaled by velocity scale of $U_w$ as mentioned in section~\ref{sec:nondimensionalization}.
The plot data uses the maximum slip velocity and the instantaneous value of fluid shear rate and gradient of shear rate at the wall corresponding to the occurrence of maximum slip. A universal curve is well described by the form $y={a}/{x^{b}}$ where $y=\tilde{U_s}$ and $x=C_{nd}\frac {\partial^2 U_f^*}{\partial y^{* 2}}  + \frac{\partial U_f^*}{\partial y^*}$ and the dashed line represents the value $a=1.0$ and $b=1.0$. The exact fitting coefficients to our experimental data were calculated to be $a=1.05$ and $b=1.14$.

Proper scaling leads to the collapse of data which suggests a universal boundary condition for any wall-fluid property. The universal curve equips us with a very potent tool in determining the slip at the solid-fluid boundary for a unsteady and steady flows having the knowledge of fluid shear rate. Thereby eliminating the need to perform extensive molecular dynamic simulations which are computationally expensive.

\subsection{Hysteresis}
\begin{figure}
  \subfigure[~]{\label{fig:Hyst_1}\includegraphics[width=0.43\textwidth]{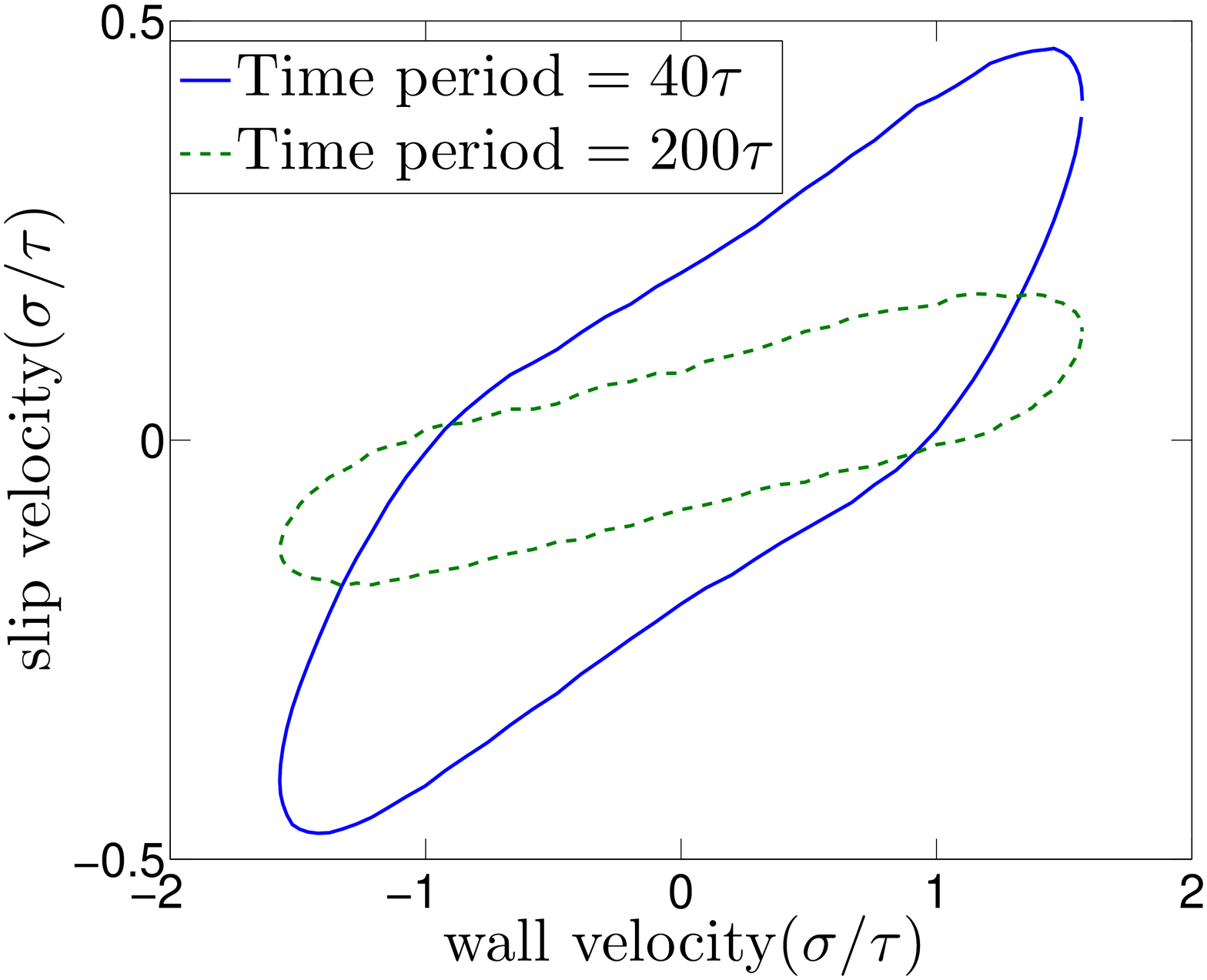}}
  \subfigure[~]{\label{fig:Hyst_2}\includegraphics[width=0.43\textwidth]{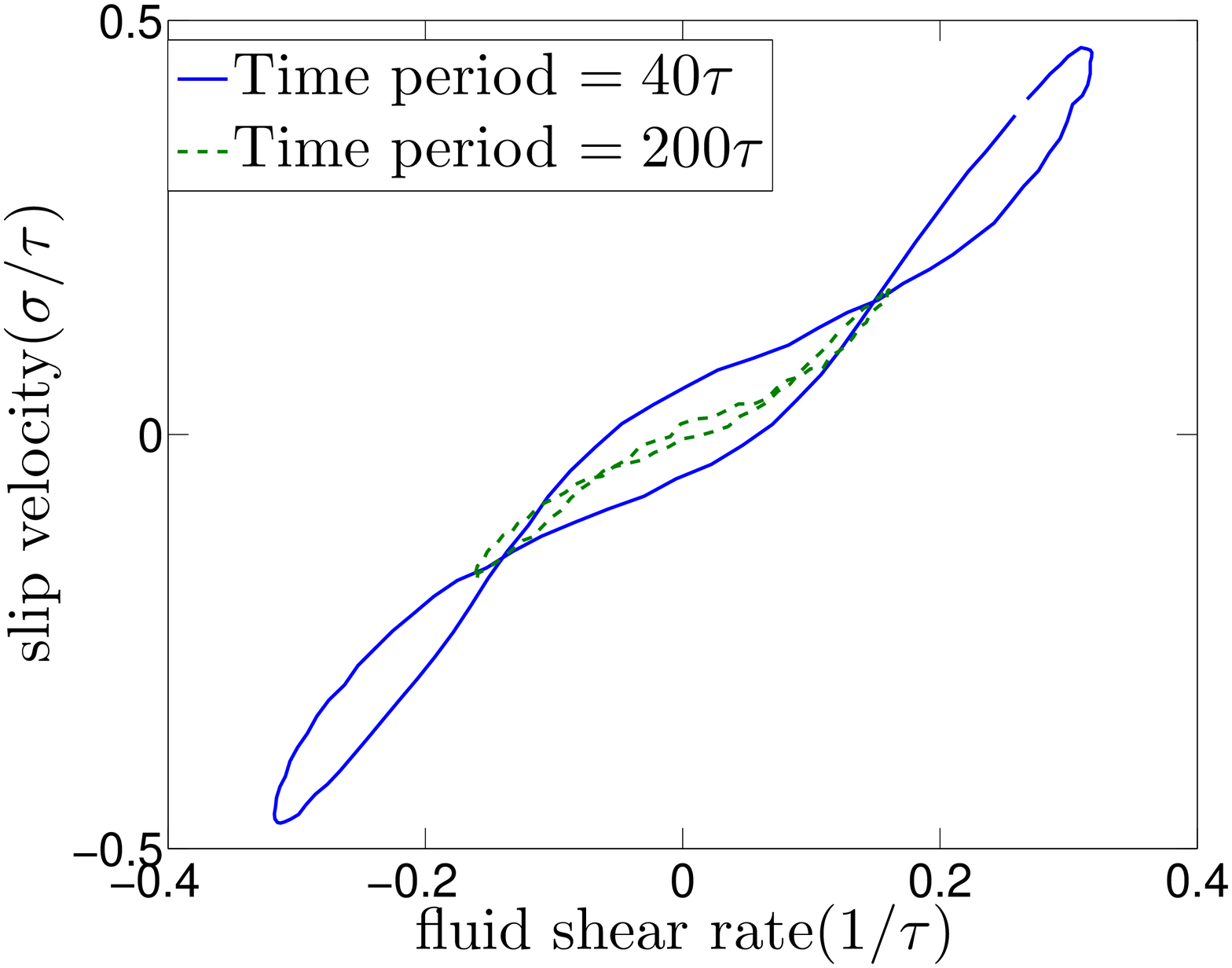}}

 \caption{Slip velocity as a function of (a) wall velocity and (b) fluid shear rate exhibiting hysteresis is shown. The plots are for case 1 and an amplitude of $10\sigma$ and $50\sigma$ corresponding to a time period of $40\tau$ and $200\tau$ was chosen in order to maintain the same maximum wall velocity. All the curves are for a complete cycle of oscillations. It is seen that the area confined by the hysteresis loop decreases with increase in time period.
}
 \label{fig:Hyst}
\end{figure}

The lagging of fluid velocity due to fluid inertia suggests the presence of hysteresis and forms the motivation to explore it in unsteady oscillating flows. In Fig.~\ref{fig:Hyst}(a) the slip velocity at the wall is plotted against the wall velocity and in Fig.~\ref{fig:Hyst}(b) change in slip velocity is plotted against the fluid shear rate. Two different time period of wall oscillation, $40\tau$ and $200\tau$, are considered in each figure to study the effects of inertia on hysteresis. A hysteresis loop is formed in both the figures and the area confined within the loop is seen to decrease with increasing time period.

The effect of inertia in developing hysteresis as seen in Fig.~\ref{fig:Hyst} is explained here. This could be described by visualizing fluid layers in the vicinity of the wall during the time when the wall is approaching an extreme location in the cycle and then changing direction. As the wall approaches the extreme point in the oscillation all the fluid atoms are moving in the same direction as the wall. But, when the wall changes its direction the fluid adjacent to the wall does not change direction instantaneously as a result of inertia of fluid atoms in this and the neighboring layers. This leads to a change in the magnitude of fluid velocity, depending on the direction of motion of wall, although they are evaluated at the same location of wall in the cycle.

While there is a distinct loop observed in Fig.~\ref{fig:Hyst}(a) at a time period of $200\tau$ the loop in Fig.~\ref{fig:Hyst}(b) collapses into a single curve. The area formed by slip velocity and wall velocity has the dimensions of energy per unit mass while that formed by slip velocity and fluid shear rate has the dimensions of acceleration. Hence, the area in Fig.~\ref{fig:Hyst}(a) could be possibly interpreted as the additional amount of energy the wall needs to transfer to the fluid below it in order to attain thermal equilibrium. This can be related to the lack of thermal equilibrium of fluid atoms adjacent to the wall which leads to slip. The area in Fig.~\ref{fig:Hyst}(b) having the dimensions of acceleration can be correlated to inertia in the fluid. As the time period is increased from $40\tau$ to $200\tau$ we essentially reduce the inertia in the flow thereby getting closer to steady state which leads to the collapse of the loop.

The hysteresis here can be seen to be analogous to the commonly known magnetic hysteresis~\cite{BertottiG:98a}. When a ferromagnetic material is subjected to magnetic field the atomic dipoles aligns itself with the field but when the field is reduced to zero, partial alignment is retained. In order to demagnetize it a field must be applied in opposite direction. This results in the difference in the path taken by the two legs in the hysteresis loop.

\section{Conclusion}
\label{sec:conclusion}
A series of numerical experiments with different wall-fluid interaction properties and varying amplitudes and time periods were performed for a time-periodic oscillatory Couette flow problem. An increase in slip is observed as compared to the steady Couette flow problem. This increased slip was attributed to fluid inertial forces not represented in a steady flow. A case having wall-fluid properties that showed no-slip for Thompson and Troian's steady Couette flow experiment is chosen. For this case when the fluid is subjected to an oscillatory flow a distinct slip is observed which confirms the increase in slip in an unsteady flow. To gain a deeper insight into the cause of the increased slip an unsteady slip model is established based on the Maxwell's slip boundary condition. The dependence of slip on acceleration in addition to shear rate is shown. By writing acceleration in terms of shear rate, it is shown that slip at the wall depends on gradient of shear rate and the shear rate of fluid at the wall. For the limiting case of steady flow the model reduces back to the Maxwell's model. This provides a more accurate prediction of slip for unsteady flow problem rather than simply using the steady Navier's or Maxwell's slip model. Non-dimensionalizing the model by scaling the problem by the wall velocity and the characteristic length of Stokes second problem leads to the collapse of data onto a single universal curve. Thereby with the knowledge of shear rate of fluid, slip length can be calculated using the universal curve without having to perform any computationally expensive MD simulations. A key non-dimensional number, defined as the ratio of phase speed to speed of sound, helps in explaining and characterizing the transition of slip boundary condition from finite to perfect slip is also identified from this non-dimensionalization. Phase lag in fluid velocity relative to wall is observed. The lag increases with decreasing time period of wall oscillation and increasing hydrophobicity.

 Hysteresis is observed while comparing slip velocity with wall velocity and shear rate. The cause for hysteresis can be attributed to the inertia of fluid. It is seen that the area formed by the loop decreases with increase in time period of wall oscillation. For the case of slip velocity versus the shear rate the loop collapses to a curve as the flow tends to a steady flow at an increased time period. In the case of slip velocity versus the wall velocity the area inside the hysteresis loop can be related to the loss of energy transfer from the wall to the fluid. For the loop formed by slip velocity and shear rate the area can be said to be equivalent to inertia of the flow.

\section*{Acknowledgment}
This research was supported by the Office of Naval Research.


\begin{thebibliography}{10}%
\makeatletter
\providecommand \@ifxundefined [1]{%
 \ifx #1\undefined \expandafter \@firstoftwo
 \else \expandafter \@secondoftwo
\fi
}%
\providecommand \@ifnum [1]{%
 \ifnum #1\expandafter \@firstoftwo
 \else \expandafter \@secondoftwo
\fi
}%
\providecommand \enquote [1]{``#1''}%
\providecommand \bibnamefont  [1]{#1}%
\providecommand \bibfnamefont [1]{#1}%
\providecommand \citenamefont [1]{#1}%
\providecommand\href[0]{\@sanitize\@href}%
\providecommand\@href[1]{\endgroup\@@startlink{#1}\endgroup\@@href}%
\providecommand\@@href[1]{#1\@@endlink}%
\providecommand \@sanitize [0]{\begingroup\catcode`\&12\catcode`\#12\relax}%
\@ifxundefined \pdfoutput {\@firstoftwo}{%
 \@ifnum{\z@=\pdfoutput}{\@firstoftwo}{\@secondoftwo}%
}{%
 \providecommand\@@startlink[1]{\leavevmode}%
 \providecommand\@@endlink[0]{}%
}{%
 \providecommand\@@startlink[1]{%
  \leavevmode
  \pdfstartlink
   attr{/Border[0 0 1 ]/H/I/C[0 1 1]}%
   user{/Subtype/Link/A<</Type/Action/S/URI/URI(#1)>>}%
  \relax
 }%
 \providecommand\@@endlink[0]{\pdfendlink}%
}%
\providecommand \url  [0]{\begingroup\@sanitize \@url }%
\providecommand \@url [1]{\endgroup\@href {#1}{\urlprefix}}%
\providecommand \urlprefix [0]{URL }%
\providecommand \Eprint[0]{\href }%
\@ifxundefined \urlstyle {%
  \providecommand \doi [1]{doi:\discretionary{}{}{}#1}%
}{%
  \providecommand \doi [0]{doi:\discretionary{}{}{}\begingroup
  \urlstyle{rm}\Url }%
}%
\providecommand \doibase [0]{http://dx.doi.org/}%
\providecommand \Doi[1]{\href{\doibase#1}}%
\providecommand \bibAnnote [3]{%
  \BibitemShut{#1}%
  \begin{quotation}\noindent
    \textsc{Key:}\ #2\\\textsc{Annotation:}\ #3%
  \end{quotation}%
}%
\providecommand \bibAnnoteFile [2]{%
  \IfFileExists{#2}{\bibAnnote {#1} {#2} {\input{#2}}}{}%
}%
\providecommand \typeout [0]{\immediate \write \m@ne }%
\providecommand \selectlanguage [0]{\@gobble}%
\providecommand \bibinfo [0]{\@secondoftwo}%
\providecommand \bibfield [0]{\@secondoftwo}%
\providecommand \translation [1]{[#1]}%
\providecommand \BibitemOpen[0]{}%
\providecommand \bibitemStop [0]{}%
\providecommand \bibitemNoStop [0]{.\EOS\space}%
\providecommand \EOS [0]{\spacefactor3000\relax}%
\providecommand \BibitemShut [1]{\csname bibitem#1\endcsname}%
\bibitem{GoldsteinS:38a}%
  \BibitemOpen
  \bibfield{author}{%
  \bibinfo {author} {\bibfnamefont{S.}~\bibnamefont{Goldstein}},\ }%
  \enquote{\bibinfo {title} {Modern development in fluid dynamics},}\ \
  (\bibinfo {publisher} {Clarendon Press Oxford},\ \bibinfo {year} {1938})\
  pp.\ \bibinfo {pages} {676--680}%
  \bibAnnoteFile{NoStop}{GoldsteinS:38a}%
\bibitem{GoldsteinS:69a}%
  \BibitemOpen
  \bibfield{author}{%
  \bibinfo {author} {\bibfnamefont{S.}~\bibnamefont{Goldstein}},\ }%
  \bibfield{journal}{%
  \bibinfo {journal} {Ann. Rev. Fluid Mech.}\ }%
  \textbf{\bibinfo {volume} {1}},\ \bibinfo {pages} {1} (\bibinfo {year}
  {1969})%
  \bibAnnoteFile{NoStop}{GoldsteinS:69a}%
\bibitem{VinogradovaOI:99a}%
  \BibitemOpen
  \bibfield{author}{%
  \bibinfo {author} {\bibfnamefont{O.}~\bibnamefont{Vinogradova}},\ }%
  \bibfield{journal}{%
  \bibinfo {journal} {Int. J. Miner. Process}\ }%
  \textbf{\bibinfo {volume} {56}},\ \bibinfo {pages} {31} (\bibinfo {year}
  {1999})%
  \bibAnnoteFile{NoStop}{VinogradovaOI:99a}%
\bibitem{NavierCLMH:1823a}%
  \BibitemOpen
  \bibfield{author}{%
  \bibinfo {author} {\bibfnamefont{C.}~\bibnamefont{Navier}},\ }%
  \bibfield{journal}{%
  \bibinfo {journal} {Memoires de l Academie Royale des Sciences de l
  Instituede France}\ }%
  \textbf{\bibinfo {volume} {6}},\ \bibinfo {pages} {389} (\bibinfo {year}
  {1823})%
  \bibAnnoteFile{NoStop}{NavierCLMH:1823a}%
\bibitem{MaxwellJC:90a}%
  \BibitemOpen
  \bibfield{author}{%
  \bibinfo {author} {\bibfnamefont{J.}~\bibnamefont{Maxwell}}\ }%
  (\bibinfo {year} {1890})\ pp.\ \bibinfo {pages} {703--711}%
  \bibAnnoteFile{NoStop}{MaxwellJC:90a}%
\bibitem{KoplikJ:88a}%
  \BibitemOpen
  \bibfield{author}{%
  \bibinfo {author} {\bibfnamefont{J.}~\bibnamefont{Koplik}}, \bibinfo {author}
  {\bibfnamefont{J.}~\bibnamefont{Banavar}},\ and\ \bibinfo {author}
  {\bibfnamefont{J.~F.}\ \bibnamefont{Willemsen}},\ }%
  \bibfield{journal}{%
  \Doi{10.1103/PhysRevLett.60.1282}{\bibinfo {journal} {Phys. Rev. Lett.}}\ }%
  \textbf{\bibinfo {volume} {60}},\ \bibinfo {pages} {1282} (\bibinfo {year}
  {1988})%
  \bibAnnoteFile{NoStop}{KoplikJ:88a}%
\bibitem{KoplikJ:89a}%
  \BibitemOpen
  \bibfield{author}{%
  \bibinfo {author} {\bibfnamefont{J.}~\bibnamefont{Koplik}}, \bibinfo {author}
  {\bibfnamefont{J.}~\bibnamefont{Banavar}},\ and\ \bibinfo {author}
  {\bibfnamefont{J.~F.}\ \bibnamefont{Willemsen}},\ }%
  \bibfield{journal}{%
  \bibinfo {journal} {Physics of Fluids, A}\ }%
  \textbf{\bibinfo {volume} {1}},\ \bibinfo {pages} {781} (\bibinfo {year}
  {1989})%
  \bibAnnoteFile{NoStop}{KoplikJ:89a}%
\bibitem{ThompsonP:90a}%
  \BibitemOpen
  \bibfield{author}{%
  \bibinfo {author} {\bibfnamefont{P.~A.}\ \bibnamefont{Thompson}}\ and\
  \bibinfo {author} {\bibfnamefont{M.}~\bibnamefont{Robbins}},\ }%
  \bibfield{journal}{%
  \bibinfo {journal} {\PRA}\ }%
  \textbf{\bibinfo {volume} {41}} (\bibinfo {year} {1990}),\ \doi{\bibinfo
  {doi} {10.1103/PhysRevA.41.6830}}%
  \bibAnnoteFile{NoStop}{ThompsonP:90a}%
\bibitem{TroianSM:97a}%
  \BibitemOpen
  \bibfield{author}{%
  \bibinfo {author} {\bibfnamefont{P.~A.}\ \bibnamefont{Thompson}}\ and\
  \bibinfo {author} {\bibfnamefont{S.~M.}\ \bibnamefont{Troian}},\ }%
  \bibfield{journal}{%
  \bibinfo {journal} {Nature}\ }%
  \textbf{\bibinfo {volume} {389}},\ \bibinfo {pages} {360} (\bibinfo {month}
  {25 September}\ \bibinfo {year} {1997})%
  \bibAnnoteFile{NoStop}{TroianSM:97a}%
\bibitem{BarratJ:99a}%
  \BibitemOpen
  \bibfield{author}{%
  \bibinfo {author} {\bibfnamefont{J.}~\bibnamefont{Barrat}}\ and\ \bibinfo
  {author} {\bibfnamefont{L.}~\bibnamefont{Bocquet}},\ }%
  \bibfield{journal}{%
  \bibinfo {journal} {Phys. Rev. Lett.}\ }%
  \textbf{\bibinfo {volume} {82}},\ \bibinfo {pages} {4671} (\bibinfo {year}
  {1999})%
  \bibAnnoteFile{NoStop}{BarratJ:99a}%
\bibitem{LandmanU:00a}%
  \BibitemOpen
  \bibfield{author}{%
  \bibinfo {author} {\bibfnamefont{J.}~\bibnamefont{Gao}}, \bibinfo {author}
  {\bibfnamefont{W.}~\bibnamefont{Luedtke}},\ and\ \bibinfo {author}
  {\bibfnamefont{U.}~\bibnamefont{Landman}},\ }%
  \bibfield{journal}{%
  \bibinfo {journal} {Tribology Lett.}\ }%
  \textbf{\bibinfo {volume} {9}},\ \bibinfo {pages} {3} (\bibinfo {year}
  {2000})%
  \bibAnnoteFile{NoStop}{LandmanU:00a}%
\bibitem{RobbinsMO:01a}%
  \BibitemOpen
  \bibfield{author}{%
  \bibinfo {author} {\bibfnamefont{C.}~\bibnamefont{Denniston}}\ and\ \bibinfo
  {author} {\bibfnamefont{M.}~\bibnamefont{Robbins}},\ }%
  \bibfield{journal}{%
  \bibinfo {journal} {Phys. Rev. Lett.}\ }%
  \textbf{\bibinfo {volume} {87}} (\bibinfo {year} {2001})%
  \bibAnnoteFile{NoStop}{RobbinsMO:01a}%
\bibitem{TroianSM:04a}%
  \BibitemOpen
  \bibfield{author}{%
  \bibinfo {author} {\bibfnamefont{N.~V.}\ \bibnamefont{Priezjev}}\ and\
  \bibinfo {author} {\bibfnamefont{S.~M.}\ \bibnamefont{Troian}},\ }%
  \bibfield{journal}{%
  \bibinfo {journal} {\PRL}\ }%
  \textbf{\bibinfo {volume} {92}},\ \bibinfo {pages} {018302} (\bibinfo {year}
  {2004})%
  \bibAnnoteFile{NoStop}{TroianSM:04a}%
\bibitem{PriezjevNV:07a}%
  \BibitemOpen
  \bibfield{author}{%
  \bibinfo {author} {\bibfnamefont{N.}~\bibnamefont{Priezjev}},\ }%
  \bibfield{journal}{%
  \bibinfo {journal} {The Journal of Chemical Physics}\ }%
  \textbf{\bibinfo {volume} {127}},\ \bibinfo {pages} {144708(6pp.)} (\bibinfo
  {year} {2007})%
  \bibAnnoteFile{NoStop}{PriezjevNV:07a}%
\bibitem{PriezjevNV:07b}%
  \BibitemOpen
  \bibfield{author}{%
  \bibinfo {author} {\bibfnamefont{N.}~\bibnamefont{Priezjev}},\ }%
  \bibfield{journal}{%
  \bibinfo {journal} {\PRE}\ }%
  \textbf{\bibinfo {volume} {75}},\ \bibinfo {pages} {051605 (7pp.)} (\bibinfo
  {year} {2007})%
  \bibAnnoteFile{NoStop}{PriezjevNV:07b}%
\bibitem{PriezjevNV:09a}%
  \BibitemOpen
  \bibfield{author}{%
  \bibinfo {author} {\bibfnamefont{N.}~\bibnamefont{Priezjev}},\ }%
  \bibfield{journal}{%
  \bibinfo {journal} {\PRE}\ }%
  \textbf{\bibinfo {volume} {80}},\ \bibinfo {pages} {031608 (11)} (\bibinfo
  {year} {2009})%
  \bibAnnoteFile{NoStop}{PriezjevNV:09a}%
\bibitem{PriezjevNV:10a}%
  \BibitemOpen
  \bibfield{author}{%
  \bibinfo {author} {\bibfnamefont{N.}~\bibnamefont{Priezjev}},\ }%
  \bibfield{journal}{%
  \bibinfo {journal} {\PRL}\ }%
  \textbf{\bibinfo {volume} {82}},\ \bibinfo {pages} {051603 (10pp.)} (\bibinfo
  {year} {2010})%
  \bibAnnoteFile{NoStop}{PriezjevNV:10a}%
\bibitem{IsraelachviliJN:96a}%
  \BibitemOpen
  \bibfield{author}{%
  \bibinfo {author} {\bibfnamefont{S.}~\bibnamefont{Campbell}}, \bibinfo
  {author} {\bibfnamefont{G.}~\bibnamefont{Luengo}}, \bibinfo {author}
  {\bibfnamefont{V.}~\bibnamefont{Srdanov}}, \bibinfo {author}
  {\bibfnamefont{F.}~\bibnamefont{Wudl}},\ and\ \bibinfo {author}
  {\bibfnamefont{J.}~\bibnamefont{Israelachvili}},\ }%
  \bibfield{journal}{%
  \bibinfo {journal} {Nature}\ }%
  \textbf{\bibinfo {volume} {382}},\ \bibinfo {pages} {520} (\bibinfo {year}
  {1996})%
  \bibAnnoteFile{NoStop}{IsraelachviliJN:96a}%
\bibitem{ChuraevNV:99a}%
  \BibitemOpen
  \bibfield{author}{%
  \bibinfo {author} {\bibfnamefont{O.}~\bibnamefont{Kiseleva}}, \bibinfo
  {author} {\bibfnamefont{V.}~\bibnamefont{Sobolev}},\ and\ \bibinfo {author}
  {\bibfnamefont{N.}~\bibnamefont{Churaev}},\ }%
  \bibfield{journal}{%
  \bibinfo {journal} {J. Colloid}\ }%
  \textbf{\bibinfo {volume} {61}},\ \bibinfo {pages} {263} (\bibinfo {year}
  {1999})%
  \bibAnnoteFile{NoStop}{ChuraevNV:99a}%
\bibitem{CraigV:01a}%
  \BibitemOpen
  \bibfield{author}{%
  \bibinfo {author} {\bibfnamefont{C.~N.}\ \bibnamefont{{V. Craig}}}\ and\
  \bibinfo {author} {\bibfnamefont{D.}~\bibnamefont{Williams}},\ }%
  \bibfield{journal}{%
  \bibinfo {journal} {Phys. Rev. Lett.}\ }%
  \textbf{\bibinfo {volume} {87}},\ \bibinfo {pages} {054504} (\bibinfo {year}
  {2001})%
  \bibAnnoteFile{NoStop}{CraigV:01a}%
\bibitem{ZhuY:01a}%
  \BibitemOpen
  \bibfield{author}{%
  \bibinfo {author} {\bibfnamefont{Y.}~\bibnamefont{Zhu}}\ and\ \bibinfo
  {author} {\bibfnamefont{S.}~\bibnamefont{Granick}},\ }%
  \bibfield{journal}{%
  \bibinfo {journal} {Phys. Rev. Lett.}\ }%
  \textbf{\bibinfo {volume} {87}} (\bibinfo {year} {2001})%
  \bibAnnoteFile{NoStop}{ZhuY:01a}%
\bibitem{ButtHJ:02a}%
  \BibitemOpen
  \bibfield{author}{%
  \bibinfo {author} {\bibfnamefont{E.}~\bibnamefont{Bonaccurso}}, \bibinfo
  {author} {\bibfnamefont{M.}~\bibnamefont{Kappal}},\ and\ \bibinfo {author}
  {\bibfnamefont{H.}~\bibnamefont{Butt}},\ }%
  \bibfield{journal}{%
  \bibinfo {journal} {Phys. Rev. Lett.}\ }%
  \textbf{\bibinfo {volume} {88}} (\bibinfo {year} {2002})%
  \bibAnnoteFile{NoStop}{ButtHJ:02a}%
\bibitem{MazuyerD:02a}%
  \BibitemOpen
  \bibfield{author}{%
  \bibinfo {author} {\bibfnamefont{J.}~\bibnamefont{Baudry}}, \bibinfo {author}
  {\bibfnamefont{E.}~\bibnamefont{Charalix}}, \bibinfo {author}
  {\bibfnamefont{A.}~\bibnamefont{Tonck}},\ and\ \bibinfo {author}
  {\bibfnamefont{D.}~\bibnamefont{Mazuyer}},\ }%
  \bibfield{journal}{%
  \bibinfo {journal} {Langmuir}\ }%
  \textbf{\bibinfo {volume} {17}},\ \bibinfo {pages} {5232} (\bibinfo {year}
  {2002})%
  \bibAnnoteFile{NoStop}{MazuyerD:02a}%
\bibitem{MeinhartCD:02a}%
  \BibitemOpen
  \bibfield{author}{%
  \bibinfo {author} {\bibfnamefont{D.}~\bibnamefont{Tretheway}}\ and\ \bibinfo
  {author} {\bibfnamefont{C.}~\bibnamefont{Meinhart}},\ }%
  \bibfield{journal}{%
  \bibinfo {journal} {\PF}\ }%
  \textbf{\bibinfo {volume} {14}},\ \bibinfo {pages} {L9} (\bibinfo {year}
  {2002})%
  \bibAnnoteFile{NoStop}{MeinhartCD:02a}%
\bibitem{LeeHJ:03a}%
  \BibitemOpen
  \bibfield{author}{%
  \bibinfo {author} {\bibfnamefont{S.}~\bibnamefont{Granick}}, \bibinfo
  {author} {\bibfnamefont{Y.}~\bibnamefont{Zhu}},\ and\ \bibinfo {author}
  {\bibfnamefont{H.}~\bibnamefont{Lee}},\ }%
  \bibfield{journal}{%
  \bibinfo {journal} {Nature Materials}\ }%
  \textbf{\bibinfo {volume} {2}},\ \bibinfo {pages} {221} (\bibinfo {year}
  {2003})%
  \bibAnnoteFile{NoStop}{LeeHJ:03a}%
\bibitem{HillJM:09a}%
  \BibitemOpen
  \bibfield{author}{%
  \bibinfo {author} {\bibfnamefont{M.}~\bibnamefont{Matthews}}\ and\ \bibinfo
  {author} {\bibfnamefont{J.}~\bibnamefont{Hill}},\ }%
  \bibfield{journal}{%
  \bibinfo {journal} {Microfluid Nanofluid}\ }%
  \textbf{\bibinfo {volume} {6}},\ \bibinfo {pages} {611} (\bibinfo {year}
  {2009})%
  \bibAnnoteFile{NoStop}{HillJM:09a}%
\bibitem{VafaiK:04a}%
  \BibitemOpen
  \bibfield{author}{%
  \bibinfo {author} {\bibfnamefont{A.}~\bibnamefont{Khaled}}\ and\ \bibinfo
  {author} {\bibfnamefont{K.}~\bibnamefont{Vafai}},\ }%
  \bibfield{journal}{%
  \bibinfo {journal} {International Journal of Non-Linear Mechanics}\ }%
  \textbf{\bibinfo {volume} {39}},\ \bibinfo {pages} {795} (\bibinfo {year}
  {2004})%
  \bibAnnoteFile{NoStop}{VafaiK:04a}%
\bibitem{BeskokA:04a}%
  \BibitemOpen
  \bibfield{author}{%
  \bibinfo {author} {\bibfnamefont{J.}~\bibnamefont{Park}}, \bibinfo {author}
  {\bibfnamefont{P.}~\bibnamefont{Bahukudumbi}},\ and\ \bibinfo {author}
  {\bibfnamefont{A.}~\bibnamefont{Beskok}},\ }%
  \bibfield{journal}{%
  \bibinfo {journal} {Physics of Fluids}\ }%
  \textbf{\bibinfo {volume} {16}} (\bibinfo {year} {2004})%
  \bibAnnoteFile{NoStop}{BeskokA:04a}%
\bibitem{ZhangYH:08a}%
  \BibitemOpen
  \bibfield{author}{%
  \bibinfo {author} {\bibfnamefont{G.}~\bibnamefont{Tang}}, \bibinfo {author}
  {\bibfnamefont{X.}~\bibnamefont{Gu}}, \bibinfo {author}
  {\bibfnamefont{R.}~\bibnamefont{Barber}}, \bibinfo {author}
  {\bibfnamefont{D.}~\bibnamefont{Emerson}},\ and\ \bibinfo {author}
  {\bibfnamefont{Y.}~\bibnamefont{Zhang}},\ }%
  \bibfield{journal}{%
  \bibinfo {journal} {\PRE}\ }%
  \textbf{\bibinfo {volume} {78}} (\bibinfo {year} {2008})%
  \bibAnnoteFile{NoStop}{ZhangYH:08a}%
\bibitem{WangCY:11a}%
  \BibitemOpen
  \bibfield{author}{%
  \bibinfo {author} {\bibfnamefont{C.}~\bibnamefont{Ng}}\ and\ \bibinfo
  {author} {\bibfnamefont{C.}~\bibnamefont{Wang}},\ }%
  \bibfield{journal}{%
  \bibinfo {journal} {Journal of Fluid Engineering}\ }%
  \textbf{\bibinfo {volume} {133}} (\bibinfo {year} {2011})%
  \bibAnnoteFile{NoStop}{WangCY:11a}%
\bibitem{HillJM:07a}%
  \BibitemOpen
  \bibfield{author}{%
  \bibinfo {author} {\bibfnamefont{M.}~\bibnamefont{Matthews}}\ and\ \bibinfo
  {author} {\bibfnamefont{J.}~\bibnamefont{Hill}},\ }%
  \bibfield{journal}{%
  \bibinfo {journal} {Acta Mechanica}\ }%
  \textbf{\bibinfo {volume} {191}},\ \bibinfo {pages} {195} (\bibinfo {year}
  {2007})%
  \bibAnnoteFile{NoStop}{HillJM:07a}%
\bibitem{KarniadakisG:05a}%
  \BibitemOpen
  \bibfield{author}{%
  \bibinfo {author} {\bibfnamefont{G.}~\bibnamefont{Karniadakis}}, \bibinfo
  {author} {\bibfnamefont{A.}~\bibnamefont{Beskok}},\ and\ \bibinfo {author}
  {\bibfnamefont{N.}~\bibnamefont{Aluru}},\ }%
  \emph{\bibinfo {title} {Microflows and Nanoflows, Fundamentals and
  Simulation}},\ Vol.~\bibinfo {volume} {29}\ (\bibinfo {publisher}
  {Springer},\ \bibinfo {year} {2005})%
  \bibAnnoteFile{NoStop}{KarniadakisG:05a}%
\bibitem{BeskokA:03a}%
  \BibitemOpen
  \bibfield{author}{%
  \bibinfo {author} {\bibfnamefont{P.}~\bibnamefont{Bahukudumbi}}, \bibinfo
  {author} {\bibfnamefont{J.}~\bibnamefont{Park}},\ and\ \bibinfo {author}
  {\bibfnamefont{A.}~\bibnamefont{Beskok}},\ }%
  \bibfield{journal}{%
  \bibinfo {journal} {Microscale Thermophysical Engineering}\ }%
  \textbf{\bibinfo {volume} {7}},\ \bibinfo {pages} {291} (\bibinfo {year}
  {2003})%
  \bibAnnoteFile{NoStop}{BeskokA:03a}%
\bibitem{PlimptonS:95a}%
  \BibitemOpen
  \bibfield{author}{%
  \bibinfo {author} {\bibfnamefont{S.}~\bibnamefont{Plimpton}},\ }%
  \bibfield{journal}{%
  \bibinfo {journal} {\JCP}\ }%
  \textbf{\bibinfo {volume} {117}},\ \bibinfo {pages} {1} (\bibinfo {year}
  {1995})%
  \bibAnnoteFile{NoStop}{PlimptonS:95a}%
\bibitem{VerletL:67a}%
  \BibitemOpen
  \bibfield{author}{%
  \bibinfo {author} {\bibfnamefont{L.}~\bibnamefont{Verlet}},\ }%
  \bibfield{journal}{%
  \bibinfo {journal} {Phys. Rev.}\ }%
  \textbf{\bibinfo {volume} {159}} (\bibinfo {month} {July}\ \bibinfo {year}
  {1967})%
  \bibAnnoteFile{NoStop}{VerletL:67a}%
\bibitem{TildesleyDJ:87a}%
  \BibitemOpen
  \bibfield{author}{%
  \bibinfo {author} {\bibfnamefont{M.}~\bibnamefont{Allen}}\ and\ \bibinfo
  {author} {\bibfnamefont{D.}~\bibnamefont{Tildesley}},\ }%
  \emph{\bibinfo {title} {Computer Simulation of Liquids}}\ (\bibinfo
  {publisher} {Clarendon Press Oxford},\ \bibinfo {address} {England},\
  \bibinfo {year} {1987})%
  \bibAnnoteFile{NoStop}{TildesleyDJ:87a}%
\bibitem{LoebLB:34a}%
  \BibitemOpen
  \bibfield{author}{%
  \bibinfo {author} {\bibfnamefont{L.}~\bibnamefont{Loeb}},\ }%
  \emph{\bibinfo {title} {The Kinetic Theory of Gases}}\ (\bibinfo {publisher}
  {McGraw-Hill Book Company},\ \bibinfo {year} {1934})%
  \bibAnnoteFile{NoStop}{LoebLB:34a}%
\bibitem{GombosiTI:94a}%
  \BibitemOpen
  \bibfield{author}{%
  \bibinfo {author} {\bibfnamefont{T.}~\bibnamefont{Gombosi}},\ }%
  \emph{\bibinfo {title} {Gaskinetic theory}}\ (\bibinfo {publisher} {Cambridge
  University Press},\ \bibinfo {year} {1994})%
  \bibAnnoteFile{NoStop}{GombosiTI:94a}%
\bibitem{FletcherR:71a}%
  \BibitemOpen
  \bibfield{author}{%
  \bibinfo {author} {\bibfnamefont{R.}~\bibnamefont{Fletcher}},\ }%
  \emph{\bibinfo {title} {A modified {M}arquardt subroutine for nonlinear least
  squares}},\ \bibinfo {type} {Tech. Rep.}\ \bibinfo {number} {R6799}\
  (\bibinfo {institution} {Atomic Energy Research Establishment},\ \bibinfo
  {address} {Harwell, England},\ \bibinfo {year} {1971})%
  \bibAnnoteFile{NoStop}{FletcherR:71a}%
\bibitem{BaldaM:07a}%
  \BibitemOpen
  \bibfield{author}{%
  \bibinfo {author} {\bibfnamefont{M.}~\bibnamefont{Balda}},\ }%
  \emph{\bibinfo {title} {{LMF}solve: Levenberg-{M}arquardt-{F}letcher’s
  algoritm for nonlinear least squares problem Balda, M.}},\ \bibinfo {type}
  {Tech. Rep.}\ (\bibinfo {institution} {MathWorks},\ \bibinfo {year} {2007})%
  \bibAnnoteFile{NoStop}{BaldaM:07a}%
\bibitem{RiceSA:60a}%
  \BibitemOpen
  \bibfield{author}{%
  \bibinfo {author} {\bibfnamefont{R.}~\bibnamefont{Harris}}\ and\ \bibinfo
  {author} {\bibfnamefont{S.}~\bibnamefont{Rice}},\ }%
  \bibfield{journal}{%
  \bibinfo {journal} {The Journal of Chemical Physics}\ }%
  \textbf{\bibinfo {volume} {33}} (\bibinfo {year} {1960})%
  \bibAnnoteFile{NoStop}{RiceSA:60a}%
\bibitem{BertottiG:98a}%
  \BibitemOpen
  \bibfield{author}{%
  \bibinfo {author} {\bibfnamefont{G.}~\bibnamefont{Bertotti}},\ }%
  \emph{\bibinfo {title} {Hysteresis in Magnetism: For Physicists, Materials
  Scientists, and Engineers}}\ (\bibinfo {publisher} {Academic Press},\
  \bibinfo {year} {1998})%
  \bibAnnoteFile{NoStop}{BertottiG:98a}%
\end{thebibliography}

%

\end{document}